\documentclass[conference]{IEEEtran}
\ifx\pdfoutput\undefined
\bibliography{mybib}
\usepackage{graphicx}
\else
\usepackage[pdftex]{graphicx}
\fi
\usepackage{url}
\usepackage{array}
\usepackage{stfloats}
\usepackage{amssymb}
\usepackage{amsmath}
\usepackage{cite}
\usepackage{enumerate}
\usepackage{epstopdf}

\begin{document}
 \title{A Multicast Approach for Constructive Interference Precoding in MISO
Downlink Channel}
 \author{
  Maha Alodeh \quad Symeon Chatzinotas \quad Bj\"{o}rn Ottersten\\
  \authorblockA{
  SnT-Interdisciplinary Centre for Security, Reliability and Trust, University of Luxembourg\\
  4, rue Alphonse Weicker, L-2721 Luxembourg\\
  e-mail:\{maha.alodeh, symeon.chatzinotas, bjorn.ottersten\}@uni.lu
               }
 } 
 
 \maketitle
 \begin{abstract}
 This paper studies the concept of jointly utilizing the data information
(DI)
and channel state information (CSI) in order to design symbol-level precoders for a multiple input and single output (MISO)
downlink channel. In this direction, the interference among the simultaneous data streams
is transformed to useful signal that can improve the signal to interference
noise ratio (SINR) of the downlink transmissions. We propose a maximum ratio
transmissions (MRT) based algorithm that jointly exploits DI and CSI to gain the benefits from these useful signals. In this context, a novel framework to minimize the power consumption
is proposed by formalizing
 the duality between 
 the constructive interference downlink channel and the multicast channels. The numerical results have
shown that the proposed schemes outperform other state of the art techniques.
\let\thefootnote\relax\footnote{This work was supported by the National Research Fund (FNR) of
Luxembourg under the AFR grant (reference 4919957) for the project Smart Resource Allocation Techniques for Satellite Cognitive Radio. 
}
\end{abstract}
\vspace{-0.1cm}
\section{Introduction}
In the last decade, multiuser MISO techniques have attracted a lot of attention due
to their capability of serving co-channel users in the same frequency and time
slots. The applications of these techniques vary according to the requested
service. The first service type is known as broadcast in which a transmitter
  has
a common message to be sent to multiple receivers. In physical layer research,  this service has been studied under the term of multicasting \cite{multicast}. Since a single data stream
is sent to all receivers, there is no need to combat the interuser interference.
In the remainder of this paper, this case will be referred to as multicast.
The second service type is known as unicast, in which a transmitter
has an individual message for a receiver. Due to the nature of the wireless medium
and the use of multiple antennas, multiple simultaneous unicast transmissions are possible in the downlink of a base station (BS). In these cases, multiple streams are
simultaneously sent, which motivates precoding techniques that mitigate the
interuser interference. In information theory terms, this
service type has been studied using the broadcast channel \cite{caire}. In
the remainder of this paper, this case will be referred to as downlink.

The concept of jointly using the DI and CSI in symbol level was originally proposed in \cite{Christos-1}-\cite{Christos}. Taking into account the data information, the cross correlations among the users' subspaces can be optimized. The contributions of this paper can be summarized in the following points:
\begin{itemize}
\item As preliminary step, we develop a new constructive interference algorithm, called
constructive interference maximum ratio transmissions (CIMRT). This technique is shown to outperform the constructive rotation zero forcing precoding (CRZF) in \cite{Christos}. 
\item We illustrate the relation between the constructive interference precoding and constrained
constellation multicast precoding and as a result we establish a suitable upperbound for constructive interference
precoding systems. To characterize the multicast performance, we derive  the minimum power that is required to satisfy the quality of service  of multicast MISO for MPSK inputs.

\item We verify the convexity of the power minimization with the additional
constraints.
Therefore, we formulate the optimal precoders for power minimization for
 constrained constellation multicast channel.
Finally, we show that there is a duality between constructive interference downlink channel
(CIDC) and constrained constellation multicast channel (CCMC). Based on
this duality, we derive the optimal precoding for constructive interference
downlink channel.
\end{itemize}

\textbf{Notation}:  We use boldface upper and lower case letters for
 matrices and column vectors, respectively. $(\cdot)^H$, $(\cdot)^*$
 stand for Hermitian transpose and conjugate of $(\cdot)$. $\mathbb{E}(\cdot)$ and $\|\cdot\|$ denote the statistical expectation and the Euclidean norm, $\otimes$ denotes the kronecker
product, and $\mathbf{A}\succeq \mathbf{0}$ is used to indicate the positive
semidefinite matrix. $\angle(\cdot)$, $|\cdot|$ are the angle and magnitude  of $(\cdot)$ respectively. $\mathcal{R}(\cdot)$, $\mathcal{I}(\cdot)$
 are the real and the imaginary part of $(\cdot)$. Finally, the vector of
 all zeros with length of $K$ is defined as $\mathbf{0}^{K\times 1}$.
\section{System and Signal Models}
\vspace{-0.1cm} 
We consider a single-cell multiple-antenna downlink scenario,
where a single BS is equipped with $nt$
transmit antennas that supports data traffic to $K$ user terminals,
each one of them is equipped with a single receiving antenna. 
We assume a quasi static block fading channel $\mathbf{h}_j\in\mathbb{C}^{1\times
nt}$ between
the BS antennas and the $j^{th}$ user. Let $\mathbf{w}_k$ be the $\mathbb{C}^{nt\times
1}$ normalized precoding vector for the user $j$. The received signal at $j^{th}$
user ${y}_j$ is given by
\vspace{-0.1cm}
\begin{eqnarray}
\label{rx_o}
{y}_j=\sqrt{p_j}\mathbf{h}_j\mathbf{w}_j d_j+\displaystyle\sum_{k\neq j}\sqrt{p_k}\mathbf{h}_j\mathbf{w}_k
d_k+z_j
\end{eqnarray}
where $z_j$ denotes the noise at $j^{th}$ receiver, which is assumed independent and identically distributed (i.d.d)  complex Gaussian distributed variable $\mathcal{CN}(0,1)$. A more compact system formulation
is obtained by stacking the received signals and the noise
components for the set of K selected users as $\mathbf{y}=
\mathbf{H}\mathbf{W}\mathbf{P}^{\frac{1}{2}}  \mathbf{d}+ \mathbf{z}$, with $\mathbf{H} = [\mathbf{h}_1,..., \mathbf{h}_K]^T \in\mathbb{C}^{K\times nt} $, $\mathbf{W}=[\mathbf{w}_1, ...,\mathbf{w}_K]\in\mathbb{C}^{nt\times K}$ as the
compound channel and precoding matrices. Notice that the transmitted signal $\mathbf{d}\in\mathbb{C}^{K\times 1}$
includes the uncorrelated data symbols $d_k(t)$ for all users with $\mathbb{E}[{|d_j|^2}] = 1$, $\mathbf{P}^{\frac{1}{2}}$
is the power allocation matrix $\mathbf{P}^{\frac{1}{2}}=diag(\sqrt{p}_1,\hdots,\sqrt{p}_K)$. The CSI and DI are available at the transmitter side.
\vspace{-0.1cm}
 \section{Constructive Interference Precoding}
 \vspace{-0.1cm}
The interference among the simultaneous spatial streams
leads to deviation of the received symbols from their detection region. However,
in some cases (e.g. M-PSK) this interference pushes the received symbols further into the correct detection region and, as a consequence it enhances the system performance. Therefore, the interference can
be classified into constructive or destructive based on whether it facilitates or deteriorates the correct detection of the received symbol.
\vspace{-0.1cm}
\subsection{Constructive Interference Definition}
Assuming both DI and CSI is available at the transmitter, the normalized created interference from the $k^{th}$ data stream on $j^{th}$ user can be formulated as:   
\vspace{-0.1cm}
\begin{equation}
\psi_{jk}=\frac{\mathbf{h}_{j}\mathbf{w}_k}{\|\mathbf{h}_{j}\|\|\mathbf{w}_k\|}.
\end{equation}
Since the adopted modulations are M-PSK ones, a definition for
constructive interference can be stated as
\begin{newtheorem}{lemma}{\textbf{Lemma}}
\begin{lemma}
\label{lemma}
For any M-PSK modulated symbol $d_k$ is said to receive constructive
interference from another simultaneously transmitted symbol $d_j$ which is
associated with $\mathbf{w}_j$ if and only if the following inequalities hold   
\begin{equation}
\label{one}
\angle{d_j}-\frac{\pi}{M}\leq \tan^{-1}\Bigg(\frac{\mathcal{I}\{\psi_{jk}d_{k}\}}{\mathcal{R}\{\psi_{jk}d_{k}\}}\Bigg)\leq \angle{d_j}+\frac{\pi}{M}
\end{equation}
\begin{equation}
\label{two}
\mathcal{R}\{{d_k}\}.\mathcal{R}\{\psi_{kj}
d_{j}\}>0, \mathcal{I}\{{d_k}\}.\mathcal{I}\{\psi_{kj}d_{j}\}>0.\\
\end{equation}
\end{lemma}
 \vspace{0.1cm}
\begin{proof}
For any M-PSK modulated symbol, the region of correct detection lies in $\phi_j\in[\angle d_j-\frac{\pi}{M},\angle d_j+\frac{\pi}{M}]$.
In order for the interference to be constructive, the received interfering signal
should lie in the region of the target symbol. For the first condition, the
${\arctan}(\cdot)$ function checks whether the received interfering signal
originating from the $d_k$$^{th}$ transmit symbol is located in
the detection region of the target symbol. However, the trigonometric
functions are not one-to-one functions. This means that it manages to check
the two quadrants which the interfering symbol may lie in. To find which
one of these quadrants is the correct one, an additional constraint is added to check the sign compatibility
of the target and received interfering signals. \smallskip  
\end{proof}

\end{newtheorem}
\vspace{-0.2cm}
\section{Constructive Interference Precoding for MISO Downlink Channels }
In the remainder of this paper, it is assumed that the transmitter is capable
of designing precoding on symbol level for each time instance utilizing both
CSI and DI, so that the created interference among the simultaneous
spatial streams is constructive.
\vspace{-0.2cm}    
\subsection{Correlation Rotation Zero Forcing Precoding (CRZF)}
\vspace{-0.1cm}
The precoder aims at minimizing the mean square error while it takes into
the account the rotated constructive interference \cite{Christos}. The optimization problem
can be formulated as
\vspace{-0.25cm}    
\begin{eqnarray}\nonumber
\mathcal{J}=\min_{\mathbf{W}}\quad\mathbb{E}\{\Vert\mathbf{R}_{\phi}\mathbf{d}-(\mathbf{H}\mathbf{W}\mathbf{d}+\mathbf{z})\Vert^2\}.
\end{eqnarray}
The solution can be easily expressed as 
\begin{eqnarray}
\vspace{-0.2cm}
\mathbf{W}_{CRZF}=\gamma\mathbf{H}^H(\mathbf{H}\mathbf{H}^H)^{-1}\mathbf{R}_{\phi}
\end{eqnarray}
where $\gamma=\sqrt{\frac{P}{tr\big(\mathbf{R}^H_{\phi}(\mathbf{H}\mathbf{H}^H)^{-1}\mathbf{R}_{\phi}\big)}}$
ensures the power normalization. The cross correlation
factor between the $j^{th}$ user's channel and transmitted $k^{th}$ data
stream can be expressed as
\vspace{-0.3cm}
\begin{eqnarray}
\rho_{jk}=\frac{\mathbf{h}_j\mathbf{h}^H_k}{\|\mathbf{h}_k\|\|\mathbf{h}_j\|}.
\end{eqnarray}
The relative phase $\phi_{ij}$ that grants the constructive simultaneous
transmissions can be expressed as 
\vspace{-0.2cm}
\begin{eqnarray}
\phi_{ij}=\angle d_j-\angle(\rho_{ij}.d_i).
\end{eqnarray}
The corresponding rotation matrix can be implemented as:
\vspace{-0.25cm}
\begin{eqnarray}
\mathbf{R}_{\phi} (j,k)=\rho_{jk}\exp(\phi_{jk}i),
\end{eqnarray}
and the received signal at $j^{th}$ user can be expressed as
\vspace{-0.25cm}
\begin{eqnarray}
\label{mas}
y_j\overset{a}{=}{\gamma}{{\|\mathbf{h}_j\|(\sum^K_{k=1}\rho_{jk}d_k)}}\overset{b}{=}{\gamma}\|\mathbf{h}_j\|(\sum^K_{k=1}\varepsilon_{jk})d,
\end{eqnarray}
where $\varepsilon_{jk}$ has the same magnitude as $\rho_{jk}$ but with different
phase, and $d:d \in \mathbb{C}^{1\times 1},|d|=1, \angle d=\theta,\theta
\in [0,2\pi] $.
By taking a look at (\ref{mas}-b), it has a multicast formulation since
it seems for each user that BS sends the same symbol for all users. 

\begin{newtheorem}{rem}{\textbf{Remark}}
\begin{rem}
It can be noted that this solution includes a zero forcing step and a correlation
step $\mathbf{R}_\phi$. The correlation step aims at making
the transmit signals constructively received at each user.
Unfortunately, this design fails when we deal with co-linear users $\rho_{jk}\rightarrow
1$. However, intuitively
having co-linear users should create more constructive interference and higher
gain should be anticipated. It can be easily concluded that the source of
this
contradiction is the zero forcing step. In an effort to overcome the problem, we propose a new precoding technique in the next section.   
\end{rem}
\end{newtheorem}
\subsection{Proposed Constructive Interference Maximum Ratio Transmission
(CIMRT)}
\vspace{-0.1cm}
The maximum ratio transmissions (MRT) are not suitable for simultaneous downlink
transmissions in MISO system due to the intolerable amount of the created interference. On the other hand, this feature makes it a good candidate for
constructive interference. The \textit{naive}
maximum ratio transmission (nMRT) can be formulated as 
\vspace{-0.1cm}
\begin{eqnarray}
\mathbf{W}_{\text{nMRT}}=\begin{bmatrix}\frac{{\mathbf{h}_1}^H}{\|\mathbf{h}_1\|}, \frac{{\mathbf{h}_2}^H}{\|\mathbf{h}_2\|},\hdots,\frac{{\mathbf{h}_K}^H}{\|\mathbf{h}_K\|}\end{bmatrix}.
\end{eqnarray}
A new look at the received signal can be viewed by exploiting the singular
value decomposition of $\mathbf{H}=\mathbf{S}\mathbf{V}\mathbf{D}$, and $\mathbf{W}_{MRT}=\mathbf{D}^H\mathbf{V}^{'}\mathbf{S}^H$ as follows
\vspace{-0.1cm}
\begin{eqnarray}
\label{svd}
\vspace{-0.2cm}
\mathbf{y}&=&\mathbf{H}\mathbf{W}\mathbf{d}={{\mathbf{S}\mathbf{V}\mathbf{D}\mathbf{D}^H\mathbf{V}^{'}}}{{\mathbf{S}^H}}\mathbf{P}^{1/2}\mathbf{d}
\end{eqnarray} 
\begin{eqnarray}
\mathbf{G}&=&\mathbf{S}\mathbf{V}\mathbf{V}^{'},\quad \mathbf{B}=\mathbf{S}^H
\end{eqnarray}
where $\mathbf{S}\in \mathbb{C}^{K\times K}$ is a unitary matrix that contains
the
left-singular vectors of
$\mathbf{H}$, the matrix $\mathbf{V}$ is an $K\times nt$ diagonal matrix with nonnegative real numbers on the diagonal, and $\mathbf{D}\in\mathbb{C}^{nt\times nt}$
contains right-singular vectors of $\mathbf{H}$. $\mathbf{V}^{'}$ is the power scaled of $\mathbf{V}$
to normalize each column in $\mathbf{W}_{MRT}$ to unit.
The received signal can be as
\vspace{-0.2cm}
\begin{eqnarray}
y_j=\|\mathbf{h}_j\|(\sum^K_{k=1}\sqrt{p_k}\rho_{jk}d_k).
\end{eqnarray}
\vspace{-0.05cm}
Utilizing the reformulation of $\mathbf{y}$ in
(\ref{svd}), the received signal can be written as 
\vspace{-0.05cm}
\begin{eqnarray}
\label{rot}
y_j=\|\mathbf{g}_j\|\sum^K_{k=1}\sqrt{p_k}\xi_{jk}d_k=\|\mathbf{g}_j\|\sum^K_{k=1}\sqrt{p_k}\xi_{jk}\exp(\theta_k)d
\end{eqnarray}
where $\mathbf{g}_j$ is the $j^{th}$ row of the matrix $\mathbf{G}$, $\xi_{jk}=\frac{\mathbf{g}_j\mathbf{b}_k}{\|\mathbf{g}_j\|}$. Since
$\mathbf{B}$ is a unitary matrix, it can have uncoupled rotations which can
grant the constructivity of interference.
Let $\mathbf{R}_{kj}$ be the rotation matrix in the
$(\mathbf{b}_k,\mathbf{b}_j)$-plane, which performs an orthogonal rotation of
the $k^{th}$ and $j^{th}$ columns of a unitary matrix while keeping
the others fixed, thus preserving unitarity. Assume without
loss of generality that $k >j $. The rotation matrix in the
$(\mathbf{b}_k,\mathbf{b}_j)$-plane can be written as
\hspace{-0.2cm}
\hspace{-1.5cm}\begin{eqnarray}
\label{rotation}
\hspace{-1.5cm}
\begin{tabular}{| p{7.8cm}| }
    \hline
   \footnotesize \textbf{A1}: Constructive Interference Rotation for CIMRT Algorithm\\
    \hline
    \vspace{-0.3cm}
    \begin{enumerate}
    \item    \footnotesize Find $\mathbf{P}$ assuming all the users have constructive interference.
    \item Find singular value decomposition for $\mathbf{H}=\mathbf{S}\mathbf{V}\mathbf{D}$.
    \item Construct $\mathbf{B}$, $\mathbf{G}$.
    \item Select ($\mathbf{b}_k$, $\mathbf{b}_j$)-plane for all users pair.  
    \item Find the optimal rotation parameters $\alpha$, $\delta$  for ($\mathbf{b}_k$,$\mathbf{b}_j$)
    considering $\mathbf{P}$ by solving (\ref{rotate}).
    \item Update $\mathbf{B}=\mathbf{B}\mathbf{R}_{kj}(\alpha,\delta)$
    \vspace{-0.4cm}
    \end{enumerate}\\
    \hline
\end{tabular}
\end{eqnarray}
where the non trivial entries appear at the intersections of
$k^{th}$ and $j^{th}$ rows and columns. Hence, any unitary matrix
$\mathbf{B}^{'}$ can be expressed using the following parameterization
\begin{eqnarray}
\mathbf{B}^{'}=\mathbf{B}\prod^K_{j=1}\prod^K_{k=j+1}\mathbf{R}_{kj}.
\end{eqnarray}
It can be seen from the structure of the matrix in (\ref{rotation}) that
rotation in the ($\mathbf{b}_k$,$\mathbf{b}_j$)-plane does not change the directions
of the remaining beamforming vectors. Therefore, it just modifies the value
of $\xi_{kk}$, and the precoder reads as
\vspace{-0.2cm}
\begin{eqnarray}
\mathbf{W}_{CIMRT}&=&\mathbf{D}^H\mathbf{V}^{'}\mathbf{B}^{'}.
\end{eqnarray}
To grant constructive interference, we need to rotate the ($\mathbf{b}_k$,$\mathbf{b}_j$)-plane by formulating the rotation as a set of non-linear equations as 
\vspace{-0.15cm}
\begin{eqnarray}\nonumber
\label{rotate}
\xi^{'}_{kk}d_k&=&\xi_{kk}\cos(\alpha) d_k-\xi_{kj}\sin(\alpha) e^{-j\delta}d_j\\
\xi^{'}_{jj}d_j&=&\xi_{jk}\sin(\alpha) e^{j\delta}d_k+\xi_{jj}\cos(\alpha) d_j
\end{eqnarray}
Since the set of non-linear equations can have different roots, the function
needs to be evaluated at the obtained root in order to find the optimal ones.
The optimal solution can be found when
solving for $\xi^{'}_{kk}=\sqrt{\xi^2_{kk}+\xi^2_{kj}}$, $\xi^{'}_{jj}=\sqrt{\xi^2_{jj}+\xi^2_{jk}}$.
Sometimes it is not feasible to solve for $\xi^{*}_{kk}$ and $\xi^{*}_{jj}$,
and
their values need to be reduced correspondingly. The proposed algorithm can be illustrated in the following table:  
\hspace{-0.4cm}\begin{center}
    \begin{tabular}{| p{7.8cm}| }
    \hline
   \footnotesize \textbf{A1}: Constructive Interference Rotation for CIMRT Algorithm\\
    \hline
    \vspace{-0.3cm}
    \begin{enumerate}
    \item    \footnotesize Find $\mathbf{P}$ assuming all the users have constructive interference.
    \item Find singular value decomposition for $\mathbf{H}=\mathbf{S}\mathbf{V}\mathbf{D}$.
    \item Construct $\mathbf{B}$, $\mathbf{G}$.
    \item Select ($\mathbf{b}_k$, $\mathbf{b}_j$)-plane for all users pair.  
    \item Find the optimal rotation parameters $\alpha$, $\delta$  for ($\mathbf{b}_k$,$\mathbf{b}_j$)
    considering $\mathbf{P}$ by solving (\ref{rotate}).
    \item Update $\mathbf{B}=\mathbf{B}\mathbf{R}_{kj}(\alpha,\delta)$
    \vspace{-0.4cm}
    \end{enumerate}\\
    \hline
\end{tabular}
\end{center}
\vspace{-0.1cm}

\vspace{-0.2cm}
\section{Multicast MISO systems}
\vspace{-0.1cm}
The multicast channel is defined as the channel in which a multiantenna transmitter sends a single message to multiple single antenna users \cite{multicast}-\cite{jorswieck}.
\subsection{Constrained M-PSK Multicast Transmissions}
 \vspace{-0.1cm}
The optimal input covariance for power minimization in multicast system can
be found as a solution of the following optimization
\vspace{-0.2cm} 
\begin{eqnarray}
\label{powm1}
&\underset{\mathbf{Q}:\mathbf{Q}\succeq 0}{\min}&\quad tr(\mathbf{Q})\quad s.t.\quad\mathbf{h}_j\mathbf{Q}\mathbf{h}^H_j\geq\zeta_j\quad,\forall j\in K.
\end{eqnarray}where $\zeta_j$ is the required SNR threshold for $j^{th}$ user. 
Different solutions in the literature were proposed as in \cite{multicast}. However,
for M-PSK inputs we should design the multicast precoders so that the received
signal falls into the detection region of desired symbol. Assuming a unit-rank
solution for $\mathbf{Q}$, the
optimization problem can be written as follows
\vspace{-0.2cm} 
\begin{eqnarray}\nonumber
\label{powc}
\hspace{-0.3cm}\mathbf{w}_{\text{CMC}}(d,\mathbf{H})=&\arg\underset{\mathbf{w}}{\min}&\quad tr(\mathbf{w}\mathbf{w}^H)\\\nonumber
&s.t.&\angle(\mathbf{h}_j\mathbf{w})=\angle(d)\quad\forall j\in K\\
&\quad&\mathbf{h}_j\mathbf{w}\mathbf{w}^H\mathbf{h}^H_j\geq\zeta_j\quad
\forall j\in K.
\end{eqnarray}
Here, we have $2K$ additional constraints which limit
the performance of (\ref{powc}) in comparison to (\ref{powm1}). These constraints
 grant the reception of the symbol $d$ with the target SNR  $\zeta_{j}$ for
 the $j^{th}$ user. More flexible constraints for the detection region of
 the target symbols are discussed in \cite{maha}. The minimum transmit power in (\ref{powm1})-(\ref{powc})
 occurs when the inequality constraints are replaced by equality (i.e. all
 users should achieve their target threshold SNR). We can reformulate the constraint as
 \vspace{-0.1cm}
\begin{eqnarray}\nonumber
\label{powa1}
&\underset{\mathbf{w}}{\min}&\quad (\mathbf{w}^H\mathbf{w})\\\nonumber
&s.t.&\mathcal{I}\{\mathbf{h}_j\mathbf{\mathbf{w}}\}=\sqrt{\zeta_j}\mathcal{I}\{d\}\quad, \forall j\in K\\
&\quad&\mathcal{R}\{\mathbf{h}_j\mathbf{\mathbf{w}}\}=\sqrt{\zeta_j}\mathcal{R}\{d\}\quad, \forall j\in K
\end{eqnarray}
A final formulation can expressed as 
\vspace{-0.2cm} 
\begin{eqnarray}\nonumber
\label{powa}
&\underset{\mathbf{w}}{\min}&\quad (\mathbf{w}^H\mathbf{w})\\\nonumber
&s.t.&\frac{\mathbf{h}_j\mathbf{\mathbf{w}}-(\mathbf{h}_j\mathbf{\mathbf{w}})^H}{2i}=\sqrt{\zeta_j}\mathcal{I}\{d\},
\forall
j\in K\\
&\quad&\frac{\mathbf{h}_j\mathbf{\mathbf{w}}+(\mathbf{h}_j\mathbf{\mathbf{w}})^H}{2}=\sqrt{\zeta_j}\mathcal{R}\{d\},\forall
j\in K.
\end{eqnarray}
It can be viewed that the constraints in (\ref{powc}) are turned from inequality
constraints to equality constraint (\ref{powa1})-(\ref{powa}) due to signal aligning requirements. The Lagrangian function can be derived as follows
\vspace{-0.2cm}
\begin{eqnarray}\nonumber
\hspace{-0.2cm}\mathcal{L}(\mathbf{w})&=&\mathbf{w}^H\mathbf{w}+\sum_j{\mu_j}(-0.5i\small(\mathbf{h}_j\mathbf{w}-\mathbf{w}^H\mathbf{h}^H_j\small)-\sqrt{\zeta^i_{th}}\mathcal{I}\{d\})\\
&+&\sum_j{\alpha_j}(\small(\mathbf{h}_j\mathbf{w}+\mathbf{w}^H\mathbf{h}^H_j\small)-\sqrt{\zeta^i_{th}}\mathcal{R}\{d\})
\end{eqnarray}
where $\mu_j$ and $\alpha_j$ are the Lagrangian dual variables. The derivative for the Lagrangian function can be written as
\vspace{-0.1cm}
\begin{eqnarray}
\frac{d\mathcal{L}(\mathbf{w})}{d\mathbf{w}^*}=\mathbf{w}+0.5i\sum_j\mu_j\mathbf{h}^H_j
+0.5\sum_j\alpha_j\mathbf{h}^H_j
\end{eqnarray}
\vspace{-0.15cm}
By equating this term to zero, $\mathbf{w}$ can be written as
\vspace{-0.1cm}
\begin{eqnarray}
\label{w}
\mathbf{w}_{CMC}=-0.5i\sum^K_{j=1}\mu_j\mathbf{h}^H_j
-0.5\sum_j\alpha_j\mathbf{h}^H_j \equiv\sum^K_{j=1}\nu_j\mathbf{h}^H_j
\end{eqnarray}
where $\nu_j\in \mathbb{C}=-0.5i\mu_j-0.5\alpha_j$. The optimal values of the Lagrangian variables $\mu_j$ and $\alpha_j$ can
be found by substituting $\mathbf{w}$ in the constraints (\ref{powa}) which result in solving the simultaneous set of $2K$ equations (\ref{seto}). The final constrained
constellation multicast
precoder can be found by substituting all $\mu_j$ and $\alpha_j$ in (\ref{w}).  
\begin{figure*}[t]
\vspace{-0.2cm}
\hspace{0.2cm}
\begin{tabular}[t]{c}
\begin{minipage}{17 cm}
\vspace{-1cm}
 \begin{eqnarray}
\label{multicasteq}
\begin{array}{cccc}
\label{seto}
0.5\|\mathbf{h}_1\|(\sum_k(-\mu_k+\alpha_ki)\|\mathbf{h}_k\|\rho_{1k}&-&\sum_k(-\mu_k+\alpha_ki)\|\mathbf{h}_k\|\rho^{*}_{1k})=\sqrt{\zeta_{1}}\mathcal{I}(d)\\
0.5\|\mathbf{h}_1\|(\sum_k(-\mu_ki-\alpha_k)\|\mathbf{h}_k\|\rho_{1k}&+&\sum_k(-\mu_ki-\alpha_k)\|\mathbf{h}_k\|\rho^{*}_{1k})=\sqrt{\zeta_{1}}\mathcal{R}(d)\\
\quad&\vdots&\\
0.5\|\mathbf{h}_K\|(\sum_k(-\mu_k+\alpha_ki)\|\mathbf{h}_k\|\rho_{Kk}&-&\sum_k(-\mu_k+\alpha_ki)\|\mathbf{h}_k\|\rho^{*}_{Kk})=\sqrt{\zeta_{K}}\mathcal{I}(d)\\
0.5\|\mathbf{h}_K\|(\sum_k(-\mu_ki-\alpha_k)\|\mathbf{h}_k\|\rho_{Kk}&+&\sum_k(-\mu_ki-\alpha_k)\|\mathbf{h}_k\|\rho^{*}_{Kk})=\sqrt{\zeta_{K}}\mathcal{R}(d)\\
\end{array}
\end{eqnarray}
\end{minipage}\\
\vspace{-0.3cm}\\
\hline
\hline
\end{tabular}
\end{figure*}
 \begin{newtheorem}{cor}{\textbf{Corollary}}
 \begin{cor}
 The optimal precoding for power minimization $\mathbf{w}_{CMC}$ in CCMC must span the subspaces of each user's channel.
 \end{cor}
\end{newtheorem}
Using (\ref{w}), we can rewrite the received signal at $j^{th}$ receiver
as 
\vspace{-0.3cm} 
\begin{eqnarray}\nonumber
\hspace{-1cm}\vspace{-0.2cm} 
y_j&=&\mathbf{h}_j\mathbf{w}_{CMC}d+z_j=\mathbf{h}_j\sum^K_{k=1}\nu_k\mathbf{h}^H_kd+z_j\\\vspace{-0.3cm}\nonumber
&=&\|\mathbf{h}_j\|\sum^K_{k=1}\|\mathbf{h}_k\|\nu_k\rho_{jk}d+z_j\\\vspace{-0.3cm}
&\label{cm1}\equiv&\mathbf{h}_j\begin{bmatrix}|\nu_1|*\mathbf{h}^H_1\quad \hdots\quad|\nu_K|*\mathbf{h}^H_K\end{bmatrix}
\begin{bmatrix}d*1\angle(\nu_1)\\\nonumber
\vdots\\
d*1\angle(\nu_K)\end{bmatrix}+z_j.\\
\end{eqnarray}
From (\ref{cm1}), the constellation constrained multicast can be formulated as a constructive interference downlink channel with set of precoders $\mathbf{h}^H_1,
\hdots,\mathbf{h}^H_K$, each one of these precoder is allocated with power $|\nu_k|$ and associated with symbol $d*1\angle \nu_k$.

\vspace{-0.2cm}
\subsection{From Multicast to Constructive Interference} 
\vspace{-0.1cm}
In multicast, the cross correlations among users' channel are exploited
to aid the transmission of the data symbols. The same cross correlations are
translated to interference
in the downlink channel due to the individuality of each user's message.
However, the constructive interference techniques in MISO downlink channels
exploit these cross correlations, which raise the
question about their relation with multicast techniques.
\begin{newtheorem}{thm1}{\textbf{Theorem}} 
\begin{thm1}
The optimal precoder for CIDC 
\vspace{-0.1cm}
\begin{eqnarray}\nonumber
\label{powcd}
\hspace{-0.7cm}
\mathbf{w}_{CIDC}(\mathbf{d},\mathbf{H})=&\arg\underset{\mathbf{w}}{\min}&\quad tr(\mathbf{w}\mathbf{w}^H)\\\nonumber
&s.t.&\angle(\mathbf{h}_j\mathbf{w})=\angle(d_j)\quad\forall j\in K\\
&\quad&\mathbf{h}_j\mathbf{w}\mathbf{w}^H\mathbf{h}^H_j=\zeta_j\quad
\forall j\in K.
\end{eqnarray}
is given by $\mathbf{w}_{CMC}(d,\mathbf{A}(d_j)\mathbf{H})$ in (\ref{powc}), where $\mathbf{A}(d)$
\vspace{-0.2cm}
\begin{eqnarray}
\mathbf{A}(j,k)=\begin{cases}
\exp((\angle d-\angle d_j)i), \quad j=k\\
0,\quad j\neq k.  
\end{cases}
\end{eqnarray}
\end{thm1}
\end{newtheorem}
\begin{proof} We assume that we have the following \textit{equivalent} channel as
\vspace{-0.3cm}
\begin{eqnarray}
\mathbf{H}_e=\mathbf{A}\mathbf{H}
\end{eqnarray}
The power minimization can be rewritten by replacing $\mathbf{H}$ by its
equivalent channel $\mathbf{H}_e$ in (\ref{powc}) as
\vspace{-0.1cm}
 \begin{eqnarray}\nonumber
\label{pow1}
&\underset{\mathbf{w}}{\min}&\quad (\mathbf{w}^H_e\mathbf{w}_e)\\\nonumber
&s.t.&\angle(\mathbf{h}_{e,j}\mathbf{w})=\angle(d)\quad\forall j\in K\\
&\quad&\mathbf{h}_{e,j}\mathbf{w}\mathbf{w}^H\mathbf{h}^H_{e,j}=\zeta_j\quad
\forall j\in K.
\end{eqnarray} 
where $\mathbf{h}_{e,j}$ is the $j^{th}$ row of the $\mathbf{H}_e$. The optimal
precoder $\mathbf{w}_e$ for the equivalent channel can be expressed as
\vspace{-0.2cm}
\begin{eqnarray}
\mathbf{w}_e=\sum^K_{j=1}\nu_{e,j}\mathbf{h}^H_{e,j}.
\end{eqnarray}
Rewriting the first constraints in (\ref{pow1}) as 
\vspace{-0.2cm}
\begin{eqnarray}\nonumber
&\quad&\angle(d-d_j)\angle(\mathbf{h}_{j}\mathbf{w})=\angle(d)\\
&\equiv&\angle(\mathbf{h}_{j}\mathbf{w})=\angle(d_j) \quad\forall j\in K
\end{eqnarray}
shows the equivalence  between the constrained constellation  multicast
channel and
constructive interference downlink channel. 
\end{proof}
\begin{cor}
$K$ different M-PSK symbols can be received correctly at $K$ different users by using a single precoding
 vector $\mathbf{w}\in \mathbb{C}^{n_t\times 1}$, designed according to Theorem 1, at the BS if $K\leq n_t$.
\end{cor}
\vspace{-0.2cm}
\section{Numerical Results}

The channel between the base station
and $j^{th}$ user terminal is characterized by
$\mathbf{h}_j=\sqrt{\gamma_{\circ}}\mathbf{h}^{'}_j$
where $\mathbf{h}^{'}_j\sim\mathcal{CN}(0,1)$, and $\gamma_{\circ}$ is
the average channel power. 
In order to compare all described techniques, we used an energy efficiency metric as following 
\begin{eqnarray}
\label{metric}
\eta=\frac{\sum_j R_j}{P_{tot}}.
\end{eqnarray}
where $R_j$ is the rate achieved by the $j^{th}$ user and given by $\log_2(1+\zeta_j)$,
$\zeta_j$ is selected to satisfy the related M-PSK modulation and $P_{tot}$ is the
transmitted power by BS $tr(\mathbf{W}\mathbf{W}^H)$.  
The motivation of using this metric is the fact that CRZF and CIMRT do not
aim at minimizing the transmitted power to grant certain quality of service  by their design. Therefore, in some
instance one or more users are given rate which is higher than the target rate. However for the multicast related techniques,
the optimality holds when the target rate can be achieved exactly. For the
sake of fairness in comparison, we use the metric in (\ref{metric}). 
 
\begin{figure}[h]
\begin{center}
\includegraphics[scale=0.52]{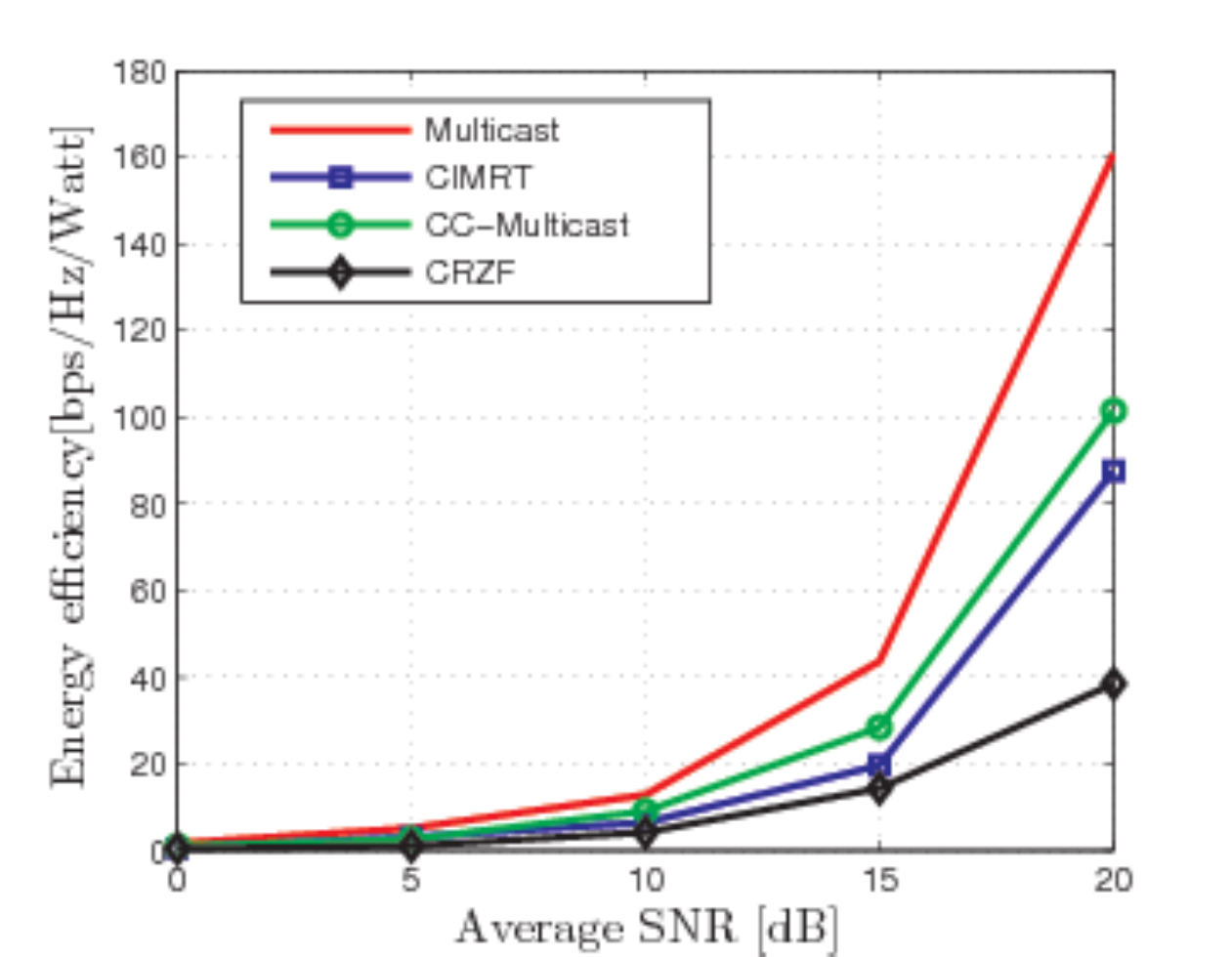}
\vspace{-0.2cm}
\caption{\label{ci}\textit{\label{fig1}\small Energy efficiency vs average SNR.}}
\end{center}
\end{figure}

\begin{figure}[h]
\begin{center}
\includegraphics[scale=0.52]{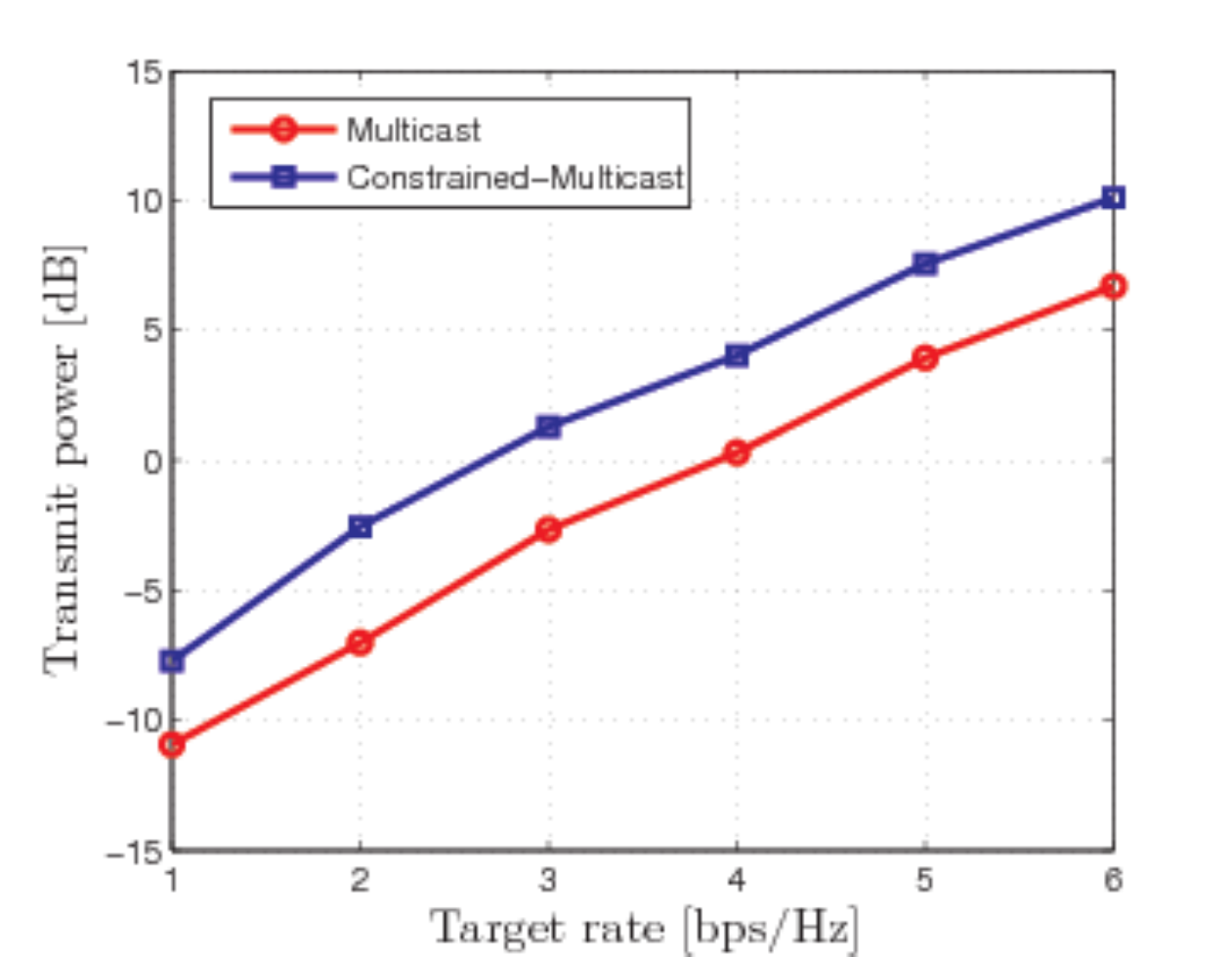}
\vspace{-0.2cm}
\caption{\label{mu}\textit{\label{fig1}\small Transmit power vs target rate.}}
\end{center}
\end{figure}

It can be seen in the Fig. (\ref{ci}) that energy efficiency for multicast with optimal
input outperforms all the proposed techniques which confirms the fact of
its optimality. For the downlink scheme derived from the constrained multicast
takes the second place.  It is shown that CRZF has the worst performance at all SNR
values. However, its peer CIMRT has a superior performance on the expense
of higher complexity due to the need for solving a non-linear set of equations.\\
Fig. (\ref{mu}) depicts the comparison between the optimal multicast technique and
constrained constellation multicast transmission from the power minimization
perspective. The considered average SNR is $10dB$, and all users
have the same target rate. It can be concluded that the optimal multicast
outperforms the constrained constellation multicast for all target rates,
due to the fact that the constrained constellation multicast requires the
phase alignment to ensure the correct  reception of the target symbols.
    
\begin{figure}[h]
\begin{center}
\includegraphics[scale=0.52]{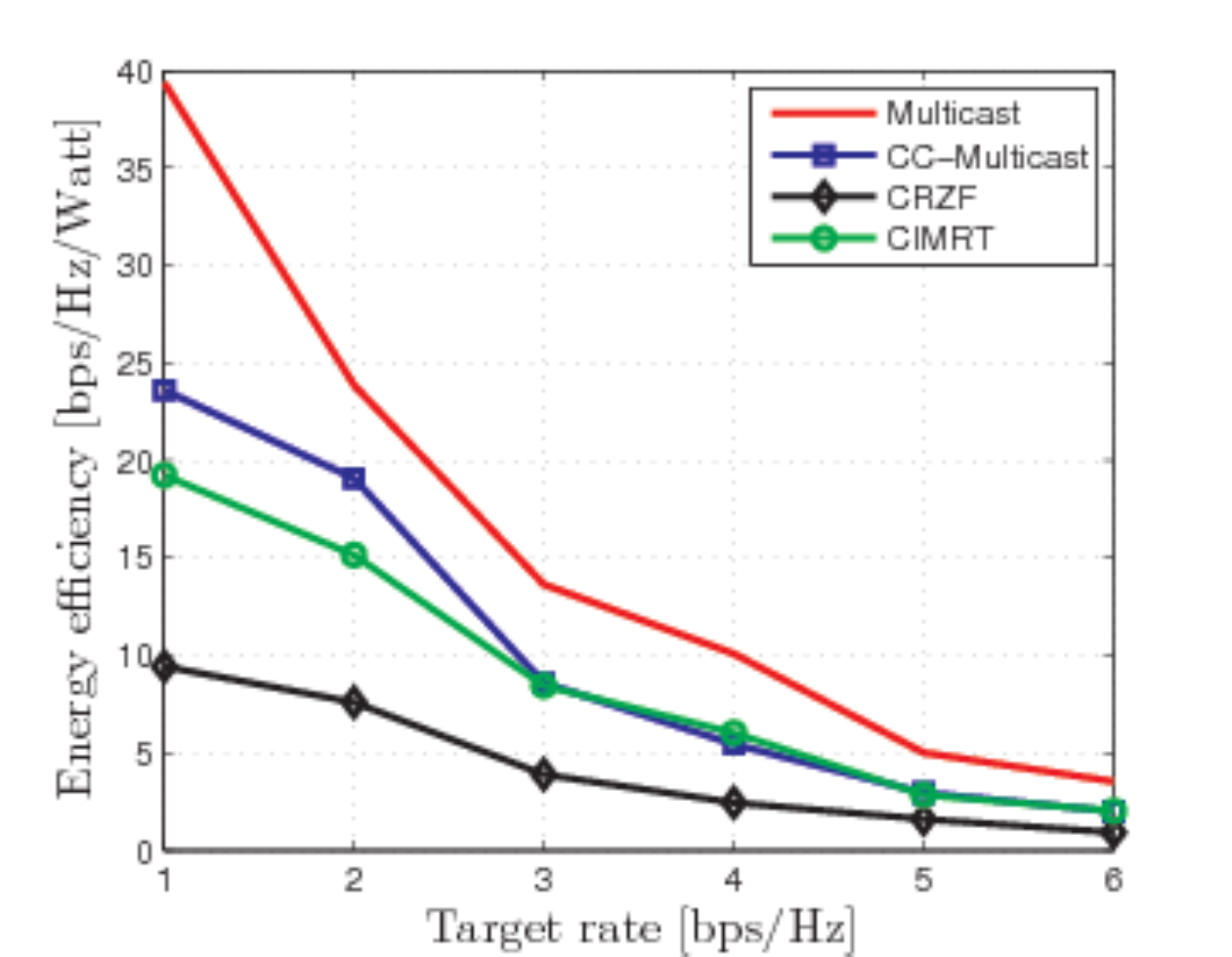}
\vspace{-0.2cm}
\caption{\label{muf5}\textit{\label{fig1}\small Energy efficiency vs target
rate}}
\end{center}
\end{figure}
Fig. (\ref{muf5}) depicts the comparison between different techniques from energy
efficiency perspective with increasing the target rates. It is clearly illustrated
that CIMRT, constrained multicast and optimal multicast  have very close
performance at high target rates.
Moreover, it can be concluded that CRZF has inferior performance with respect
to the other techniques.    

\section{conclusions}
\vspace{-0.1cm} 
In this paper, we utilized jointly CSI and DI in symbol based precoding to exploit received interfering
signal as useful energy in constructive interference precoding. In these cases, the precoding design exploits the
overlap in users' subspace instead of mitigating it. Therefore, we proposed
a new technique based on MRT to constructively correlate the interference
to enure the correct reception of data symbols. This fact enabled us
finding the connection between the constructive interference precoding and
multicast precoding wherein no interference should be mitigated. Therefore,
we found the solution for power minimization considering two inputs scenario:
the optimal input and the constrained constellation. From their closed formulations,
we concluded that their transmissions should span the subspaces of each user.
From the M-PSK constrained constellation multicast, we managed to find that the
optimal constructive
interference precoding can be expressed by solving this multicast problem
assuming an equivalent channel.

\end{document}